\documentclass[preprintnumbers, floatfix,letterpaper,aps,prd,epsfig,nofootinbib,
twocolumn
]{revtex4-1}
\usepackage{bm,graphicx,dcolumn,epstopdf,epsf, latexsym,mathbbol, amssymb,amsmath,color,slashed, mathrsfs,mathcomp,simplewick}
\usepackage[center]{subfigure}
\usepackage{multirow}
\usepackage{makecell}
\usepackage{booktabs}
\usepackage[colorlinks,linkcolor=blue,citecolor=blue,urlcolor=blue]{hyperref}

\begin{document}
\allowdisplaybreaks
 \newcommand{\bq}{\begin{equation}}
 \newcommand{\eq}{\end{equation}}
 \newcommand{\bqn}{\begin{eqnarray}}
 \newcommand{\eqn}{\end{eqnarray}}
 \newcommand{\nb}{\nonumber}
 \newcommand{\lb}{\label}
 \newcommand{\f}{\frac}
 \newcommand{\p}{\partial}
\newcommand{\PRL}{Phys. Rev. Lett.}
\newcommand{\PLB}{Phys. Lett. B}
\newcommand{\PRD}{Phys. Rev. D}
\newcommand{\CQG}{Class. Quantum Grav.}
\newcommand{\JCAP}{J. Cosmol. Astropart. Phys.}
\newcommand{\JHEP}{J. High. Energy. Phys.}
\newcommand{\bea}{\begin{eqnarray}}
\newcommand{\ena}{\end{eqnarray}}
\newcommand{\beqa}{\begin{eqnarray}}
\newcommand{\eeqa}{\end{eqnarray}}
\newcommand{\red}{\textcolor{red}}

\title{Gravitational wave constraints on non-birefringent dispersions of gravitational waves due to Lorentz violations with GWTC-3}

\author{Cheng Gong${}^{a, b, c}$}

\author{Tao Zhu${}^{b, c}$}
\email{Corresponding author: zhut05@zjut.edu.cn}

\author{Rui Niu${}^{d, e}$}

\author{Qiang Wu${}^{b, c}$}

\author{Jing-Lei Cui${}^{a}$}

\author{Xin Zhang${}^{a}$}
\email{zhangxin@mail.neu.edu.cn}

\author{Wen Zhao${}^{d, e}$}
\email{wzhao7@ustc.edu.cn}

\author{Anzhong Wang${}^{f}$}
 \email{anzhong$\_$wang@baylor.edu}

\affiliation{
${}^{a}$ Key Laboratory of Cosmology and Astrophysics (Liaoning) \& Department of Physics, College of Sciences, Northeastern University, Shenyang 110819, China\\
${}^{b}$ Institute for theoretical physics and cosmology, Zhejiang University of Technology, Hangzhou, 310032, China\\
${}^{c}$ United Center for Gravitational Wave Physics (UCGWP), Zhejiang University of Technology, Hangzhou, 310032, China\\
${}^{d}$ CAS Key Laboratory for Research in Galaxies and Cosmology, Department of Astronomy, University of Science and Technology of China, Hefei 230026, China \\
${}^{e}$ School of Astronomy and Space Sciences, University of Science and Technology of China, Hefei, 230026, China\\
${}^{f}$ GCAP-CASPER, Physics Department, Baylor University, Waco, Texas 76798-7316, USA}

\date{\today}

\begin{abstract}

The standard model extension (SME) is an effective field theory framework that can be used to study the possible violations of Lorentz symmetry in the gravitational interaction. In the SME's gauge invariant linearized gravity sector, the dispersion relation of GWs is modified, resulting in anisotropy, birefringence, and dispersion effects in the propagation of GWs. In this paper, we mainly focus on the non-birefringent and anisotropic dispersion relation in the propagation of GWs due to the violation of Lorentz symmetry. With the modified dispersion relation, we calculate the corresponding modified waveform of GWs generated by the coalescence of compact binaries. We consider the effects from the operators with the lowest mass dimension $d=6$ in the gauge invariant linearized gravity sector of the SME which are expected to have the dominant Lorentz-violating effect on the propagation of GWs. For this case, the Lorentz-violating effects are presented by 25 coefficients and we constrain them independently by the ``maximal-reach" approach. We use 90 high-confidence GW events in the GWTC-3 catalog and use {\tt Bilby}, an open source software, and {\tt Dynest}, a nested sampling package, to perform parameter estimation with the modified waveform. We do not find any evidence of Lorentz violation in the GWs data and give a $90\%$ confidence interval for each Lorentz violating coefficient. 

\end{abstract}


\maketitle

\section{Introduction}
\renewcommand{\theequation}{1.\arabic{equation}} \setcounter{equation}{0}

Classical general relativity (GR) is the most successful theory of gravity, having passed various experimental tests at different scales with astonishing accuracy \cite{Berti:2015itd, Will:1997bb, Hoyle:2000cv, Jain:2010ka, Koyama:2015vza, Clifton:2011jh, Stairs:2003eg, Manchester:2015mda, Wex:2014nva, Kramer:2016kwa}.  Although GR has been so accurate, it is not a good explanation of the theoretical singularities and quantization problems, and the experimental problems for dark matter and dark energy. On the other hand, in some candidate theories of quantum gravity, such as string theory \cite{Kostelecky:1988zi, Kostelecky:1991ak}, loop quantum gravity \cite{Gambini:1998it}, brane-worlds scenarios \cite{Burgess:2002tb}, the Lorentz invariance (LI) of the theory can be spontaneously broken.

The SME is a relatively well-developed framework for exploring Lorentz invariance violation (LIV) \cite{Colladay:1996iz, Kostelecky:2003fs}. Under this framework, all terms that may break LI can be constructed in Lagrangian. Over the past few decades, utilizing SME to test LI in the matter sector has flourished. In the gravitational sector, studies using SME to measure LI include lunar laser ranging \cite{laser1,laser2}, atom interferometers \cite{atom interferometers}, cosmic rays \cite{cosmic rays}, precision pulsar timing \cite{shao1,shao2,shao3,shao4,shao5,shao6}, planetary orbital dynamics \cite{Hees:2015mga}, and super-conducting gravimeters \cite{Flowers:2016ctv}. The case for the gravity sector of the studies is generally the coupling between gravity and matters \cite{Kostelecky:2010ze}. Since we study the Lorentz-violating effects on the propagation of GWs, we focus on the pure gravity sector with linear approximation \cite{Bailey:2006fd, Ferrari:2006gs}. We also impose the gauge-invariant condition under the usual transformation $h_{\mu \nu} \rightarrow h_{\mu \nu}+ \partial_\mu\xi_\nu + \partial_\nu\xi_\mu$, where $h_{\mu\nu}$ is the metric perturbations in the Minkovski background and $\xi_\mu$ is an arbitrary infinitesimal vector field. The discussion about gauge-violating terms can be referred to Ref. \cite{Kostelecky:2017zob}. In SME's gauge invariant linearized gravity sector, all possible general Lagrangians with quadratic metric perturbations are constructed in \cite{Kostelecky:2016kfm}, and a modified dispersion relation for GWs is also derived that can lead to anisotropy, birefringence and dispersion effects on the propagation of GWs, and it is these effects that distort the waveform of GWs. The corresponding modified waveforms of GWs are given in \cite{Mewes:2019dhj}. With the modified waveform, one can use Bayesian inference to compare the modified GWs signals with GW data and constrain the coefficients representing LIV arising from the linearized gravity sector of the SME.

The LIGO/Virgo/KAGRA detectors have detected abundant GWs sources. Now the three detectors LIGO, Virgo, and KAGRA (LVK) form the third GWs Transient Catalog (GWTC-3) \cite{LIGOScientific:2021djp}. GWTC-3 is the most comprehensive collection of GW events so far, and it plays an important role in the advancement of astrophysics \cite{LIGOScientific:2021psn}, fundamental physics, and cosmology \cite{LIGOScientific:2021aug}. GWTC-3 catalog can be used to test GR, especially the tests of Lorentz symmetry based on the phenomenological dispersion relations \cite{Will:1997bb, Mirshekari:2011yq}, which can arise from a lot of modified theories of gravity, including multifractal space-time \cite{Calcagni:2009kc}, massive gravity \cite{deRham:2014zqa, Will:1997bb}, double special relativity \cite{Amelino-Camelia:2002cqb}, Ho\v{r}ava-Lifshitz gravity \cite{Horava:2009uw, Vacaru:2012nvq} and the theory of extra dimensions \cite{Sefiedgar:2010we, Vacaru:2012nvq}. Recently, a lot of tests on the Lorentz symmetry of gravitational interaction have been carried out by using observational data from GW events in LVK catalogs. In the SME framework, the modified dispersion relation of GWs can in general lead to two other possible effects. One is the velocity birefringence of GWs which arises from the parity violation in the gravity sector of SME and causes the propagating velocities of the two polarization modes of GWs to be different. The studies on the observational effects of such velocity birefringence and their tests with signals of GW events in LVK catalogs have been performed in a lot of works \cite{Kostelecky:2016kfm, Wang:2020cub, Zhao:2019xmm, Zhao:2022pun, Wang:2020pgu, Zhu:2022uoq, Zhu:2022dfq, Niu:2022yhr, Qiao:2019hkz, Wang:2021gqm, Zhao:2019szi, Chen:2022wtz, Li:2022grj, Haegel:2022ymk, Qiao:2021fwi, ONeal-Ault:2021uwu, Sulantay:2022sag, Wang:2020pgu, Wang:2017igw}; see \cite{Qiao:2022mln} for a recent review. Another effect is the anisotropic dispersion relation which arises from the breaking of the rotation symmetry of gravity. While the tests with both the birefringence and anisotropic dispersion relation of GWs have been considered by a lot of works \cite{Shao:2020shv, Wang:2021ctl, Niu:2022yhr, Haegel:2022ymk}, the main purpose of this paper is to study the effects of the non-birefringent anisotropic dispersion relation of GWs and their observational constraints from signals of GW events in the LVK catalogs.

For our purpose, in SME's gauge invariant linearized gravity sector, we perform complete Bayesian inference using modified waveforms with effects of anisotropic non-birefringent dispersion of GWs to test the Lorentz symmetry. This paper is organized as follows. In the next section, we present a brief introduction to the propagation of GWs in the SME framework and the associated modified dispersion relation due to the effects of the LIV in the gravity sector of the SME. In Sec. \ref{sec3}, we focus on the phase modifications to the waveform of GWs due to the LIV coefficients in the SME. In Sec. \ref{sec4}, we introduce the match-filtered analysis within Bayesian inference and in Sec. \ref{sec5} we provide the constraints on the LIV coefficients by using the data of 90 GW events released in the GWTC-3 catalog. The conclusion and summary of this work are given in Sec. \ref{sec6}.

Throughout this paper, the metric convention is chosen as $(-,+,+,+)$, and greek indices $(\mu,\nu,\cdot\cdot\cdot)$ run over $0,1,2,3$ and latin indices $(i, \; j,\;k)$ run over $1, 2, 3$. We set the units to $\hbar= c = 1$.

\section{Gravitational waves in the linear gravity of SME \label{sec2}}
\renewcommand{\theequation}{2.\arabic{equation}} \setcounter{equation}{0}

In this section, we present a brief introduction to the GWs in the linearized gravity sector of the SME and the associated modified dispersion relation of GWs due to the effects of the LIV. The quadratic Lagrangian density for GWs in the gauge invariant linearized gravity sector of the SME is given by \cite{Kostelecky:2016kfm}
\bqn \lb{calL}
\mathcal{L} &=& \frac{1}{4}\epsilon^{\mu \rho \alpha \kappa} \epsilon^{\nu \sigma \beta \lambda} \eta_{\kappa \lambda} h_{\mu\nu} \partial_\alpha \partial_\beta h_{\rho \sigma} \nb\\
&&+ \frac{1}{4}h_{\mu\nu} (\hat{s}^{\mu \rho \nu \sigma} + \hat{q}^{\mu \rho \nu \sigma}+ \hat{k}^{\mu \nu \rho  \sigma}) h_{\rho \sigma},
\eqn
where one expands the metric $g_{\mu\nu}$ of the spacetime in the form of $g_{\mu \nu} = \eta_{\mu\nu} + h_{\mu\nu}$ with $\eta_{\mu\nu}$ being the constant Minkowski metric, $\epsilon^{\mu\rho \alpha \kappa}$ is the Levi-Civita tensor, and the operators $\hat{s}^{\mu \rho \nu \sigma} $, $\hat{q}^{\mu \rho \nu \sigma} $, and $\hat{k}^{\mu \rho \nu \sigma} $ represent the three different classes of modifications due to LIV. These operators can be further expanded in terms of derivatives in the following forms
\bqn
{\hat{s}}^{\mu \rho \nu \sigma} &=&\sum {s}^{(d) \mu \rho \alpha_{1} \nu \sigma \alpha_{2} \ldots \alpha_{d-2}} \partial_{\alpha_{1}} \ldots \partial_{\alpha_{d-2}}, \lb{sd}\\
{\hat{q}}^{\mu \rho \nu \sigma} &=&\sum q^{(d) \mu \rho \alpha_{1} \nu \alpha_{2} \sigma \alpha_{3} \ldots \alpha_{d-2}} \partial_{\alpha_{1}} \ldots \partial_{\alpha_{d-2}},  \lb{qd}\\
{\hat{k}}^{\mu \nu \rho \sigma}&=&\sum k^{(d) \mu \alpha_{1} \nu \alpha_{2} \rho \alpha_{3} \sigma \alpha_{4} \ldots \alpha_{d-2}} \partial_{\alpha_{1}} \ldots \partial_{\alpha_{d-2}}, \lb{kd}
\eqn
where $d$ denotes the mass dimension of the three operators and the tensor coefficients in the above expansions control the LIV. The sum in the above expansion is over even $d\geq 4$ for $s$-type violations, odd $d \geq 5$ for $q$-type, and even $d \geq 6$ for $k$-type. 
Specifically, $\hat{s}^{\mu \rho \nu \sigma}$ is anti-symmetric in both ``$\mu \rho$" and ``$\nu\sigma$", $\hat{q}^{\mu \rho \nu \sigma}$ is anti-symmetric in ``$\mu\rho$" and symmetric in ``$\nu\sigma$", and $\hat{k}^{\mu \rho \nu \sigma}$ is totally symmetric. Then the equations of motion for GWs can be derived by varying the quadratic action $S \sim \int d^4 x \mathcal{L}$ with respect to $h_{\mu\nu}$ with Lagrangian density $\mathcal{L}$ given by (\ref{calL}), which yields
\bqn\lb{eom0}
\frac{1}{2}\eta_{\rho \sigma} \epsilon^{\mu\rho\alpha \kappa} \epsilon^{\nu\sigma \beta \lambda} \partial_\alpha \partial_\beta h_{\kappa \lambda} - \delta M^{\mu\nu \rho\sigma} h_{\rho \sigma}=0,
\eqn
where the tensor operators
\bqn
\delta M^{\mu\nu\rho \sigma} &=& - \frac{1}{4} \left(\hat{s}^{\mu\rho \nu \sigma} + \hat{s}^{\mu \sigma \nu \rho}\right) - \frac{1}{2} \hat{k}^{\mu \nu \rho \sigma} \nb\\
&&-\frac{1}{8} \left(\hat{q}^{\mu \rho \nu \sigma} + \hat{q}^{\nu \rho \mu \sigma} +\hat{q}^{\mu \sigma \nu \rho} + \hat{q}^{\nu \sigma \mu \rho}\right).\nb\\
\eqn

In GR, the metric perturbation $h_{\mu \nu}$ only contains two degenerate traceless and transverse tensor modes. However, when the Lorentz-violating modifications are included, depending on specific types of the LIV, $h_{\mu\nu}$ may contain two additional scalar and two vector perturbation modes. In this paper, assuming these extra modes are small and following the similar treatment in \cite{Kostelecky:2016kfm}, we only focus on the LIV effects on the two traceless and transverse tensor modes. For this purpose, we restrict to the modes $h_{ij}$ which satisfy
\bqn
\eta^{ij} h_{ij} = 0 = \partial^i h_{ij}.
\eqn
Then the equations of motion for GWs (\ref{eom0}) reduce to
\bqn\lb{eom1}
(\partial_t^2 - \nabla^2 ) h^{ij} + 2 \delta M^{ij mn} h_{mn}=0.
\eqn
In the linearized gravity sector of SME, it is convenient to decompose the GWs into circular polarization modes. To study the evolution of $h_{ij}$, we expand it over spatial Fourier harmonics,
\bqn
h_{ij}(\tau, x^i) = \sum_{A={\rm R, L}} \int \frac{d^3 k}{(2\pi)^3}  h_{\rm A}(\tau, k^i)e^{i k_i x^i} e_{ij}^{\rm A}(k^i),\nb\\
\eqn
where $e_{ij}^{\rm A}$ denote the circular polarization tensors and satisfy the relation {$\epsilon^{i j k} n_j e_{kl}^A = i \rho_A e^{i A}_{~l}$} with $\rho_{\rm R}=1$ and $\rho_{\rm L} =-1$. So, the equations of motion in Eq.(\ref{eom1}) can be written as
\bqn
\ddot h_A + k^2 h_A + 2 \epsilon^A_{ij} \delta M^{ijmn} e^B_{mn} h_B =0,
\eqn
or equivalently in the matrix form
\begin{widetext}
\bqn
\left(
\begin{array}{ccc}
    \partial_t^2 + k^2 + 2 e_{ij}^{R} \delta M^{ijmn} e^{R}_{mn} &  2 e_{ij}^{R} \delta M^{ijmn} e^{L}_{mn} \\
2  e_{ij}^{L} \delta M^{ijmn} e^{R}_{mn}  &  \partial_t^2 + k^2 + 2 e_{ij}^{L} \delta M^{ijmn} e^{L}_{mn}
\end{array}
\right) 
\left(
\begin{array}{ccc}
   h_R  \\
h_L
\end{array}
\right) =0.
\eqn
\end{widetext}
Then, following methods developed for the study of Lorentz violation in the photon sector of the SME, the modified dispersion relation of GWs with 4-momentum $k^{\mu} = (\omega, {\bf k})$ can be derived by requiring the determinant of the above $2\times 2$ matrix vanishes, which yields (see also in \cite{Kostelecky:2016kfm})
\bqn
\omega = \Big(1- \zeta^0 \pm |{\bf \zeta }|\Big) |{\bf k}|,\lb{MD}
\eqn
where
\bqn
\zeta^0 =  - \frac{1}{2 |{\bf k}|^2} \Big( e_{ij}^{R} \delta M^{ijmn} e^{R}_{mn} + e_{ij}^{L} \delta M^{ijmn} e^{L}_{mn}\Big),\nb\\
\eqn
and
\bqn
|{\bf \zeta }|^2 &=&  \frac{1}{4 |{\bf k}|^4} \Big[ (e_{ij}^{R} \delta M^{ijmn} e^{R}_{mn} - e_{ij}^{L} \delta M^{ijmn} e^{L}_{mn})^2 \nb\\
&&+ 4 (e_{ij}^{R} \delta M^{ijmn} e^{L}_{mn}) (e_{kl}^{L} \delta M^{klpq} e^{R}_{pq})\Big].
\eqn
The modified dispersion relation in the above leads to the phase velocities of the GWs
\bqn
v_{\pm} = 1- \zeta^0 \pm |{\bf \zeta }|.
\eqn

The new effects in the modified dispersion arising from LIV are induced by the coefficients $\zeta^0$ and $|\zeta|$. While $\zeta^0$ modifies the speed of the two tensorial modes in the same way, the coefficient $|\zeta|$ leads to the two different velocities of the two tensorial modes of GWs. Therefore, the two tensorial modes can be decomposed into a fast mode (denoted by $h_f$ with velocity $v_{+}$) and a slow mode (denoted by $h_s$ with velocity $v_{-}$). This phenomenon is also known as the velocity birefringence in the propagation of GWs. It is worth noting here that the observational constraints on velocity birefringence with GW data have been extensively studied in \cite{Niu:2022yhr, Shao:2020shv, Wang:2021ctl, Niu:2022yhr, Haegel:2022ymk}. Thus in this paper, we will not perform analysis on that case and only concentrate on the effects of non-birefringent dispersion relation induced by the coefficient $\zeta^0$. 

In the modified dispersion relation (\ref{MD}), the coefficient $\zeta^0$ which leads to the non-birefringent dispersions is functions of the frequency $\omega$ and wave vector ${\bf k}$ \cite{Niu:2022yhr}. Considering it's also direction-dependent and to describe its effects on the propagation of GWs, it is convenient to expand its coefficients in terms of spin-weighted spherical harmonics $Y_{jm}$ as
\bqn
\zeta^0 &=& \sum_{d, jm} \omega^{d-4} Y_{jm}({\bf \hat n}) k_{(I)jm}^{(d)},
\eqn
where ${\bf n} = - {\bf k}$ is the direction of the source, and $0 \leq j \leq d-2$. The spherical coefficients for LIV $k_{(I)jm}^{(d)}$ are linear combinations of the tensor coefficients in (\ref{sd}, \ref{qd}, \ref{kd}). The expansions of the coefficient $\zeta^0$ are also a combination of operators at multiple mass dimensions. In general, one expects the operators with the lowest mass dimension to have the dominant Lorentz-violating effects on the propagation of GWs. In this paper, we only consider the operators with the lowest mass dimension $d$ and introduce several energy-independent coefficients as
\bqn
\zeta^0_{(d)} ({\bf n}) &=& \sum_{jm}Y_{jm}({\bf \hat n}) k_{(I)jm}^{(d)}. \lb{zeta}
\eqn
Then the phase velocity of the GWs can be rewritten as
\bqn
v  &=& 1 - \omega^{d-4} \zeta^0_{(d)}({\bf n}).  \lb{zeta0}
\eqn
The non-birefringent LIV effects are fully characterized by the coefficients, $k_{(I)jm}^{(d)}$. These coefficients determine the speeds of GWs and lead to frequency-dependent dispersions except in the case with $d=4$. Specifically, the coefficients $k_{(I)jm}^{(d)}$ are also direction-dependent if $j \neq 0$ and thus could induce the anisotropic effects on the propagation of the GWs. It is interesting to note that all these coefficients can provide frequency and direction-dependent phase modifications to the GWs. In the following, we are going to study the phase modifications due to these LIV coefficients in detail.

\section{Phase Modifications to the Waveform of GWs} \lb{sec3}
\renewcommand{\theequation}{3.\arabic{equation}} \setcounter{equation}{0}

In this section, we turn to derive the modified waveform of GW with non-birefringent LIV effects from the linearized gravity sector of the SME. For this purpose, we closely follow the derivation presented in \cite{Qiao:2019wsh, Zhao:2019xmm}. It is worth noting as well that the modified waveform has also been studied in \cite{Mewes:2019dhj}. Now consider a graviton emitted radially at $r=r_e$ and received at $r=0$, we have
\bqn
\frac{dr}{dt} = - \frac{1}{a} \left (1 - \zeta^0  \right).
\eqn
Integrating this equation from the emission time (when $r=r_e$) to arrival time (when $r=0$), one obtains
\bqn
r_e&=&  \int_{t_e}^{t_0} \frac{dt}{a(t)} -   \omega^{d-4} \zeta^0_{(d)}({\bf n}) \int_{t_e}^{t_0} \frac{dt}{a^{d-3}}.
\eqn
Considering gravitons emitted at two different times $t_e$ and $t_e'$, with wave numbers $k$ and $k'$, and received at corresponding arrival times $t_0$ and $t_0'$ ($r_e$ is the same for both),  then, the difference in their arrival times is given by 
\bqn
\Delta t_0 &=& (1+z) \Delta t_e - \Big(\omega^{d-4}_e - \omega'^{d-4}_e\Big) \zeta^0_{(d)}({\bf n}) \int_{t_e}^{t_0} \frac{dt}{a^{d-3}},\nb\\
\lb{time}
\eqn
where $z = 1/a(t_e) -1$ is the cosmological redshift. Let us focus on the GW signal generated by non-spinning, quasi-circular inspiral in the post-Newtonian approximation. Relative to the GW in GR, the LIV modifies the phase of GWs. In \cite{Will:1997bb, Mirshekari:2011yq}, it was proved that the difference of arrival times in (\ref{time}) induces the modification to the phase of GWs $\Psi$ in the following form,
\bqn
\Psi(f) =\Psi^{\rm GR} (f) - \delta \Psi(f, {\bf n}),
\eqn
where
\bqn
 \delta \Psi(f, {\bf n})  =  \frac{2^{d-3}}{d-3} \frac{u^{d-3}}{\mathcal{M}^{d-3}}  \zeta^0_{(d)}({\bf n})  \int_{t_e}^{t_0} \frac{dt}{a^{d-3}}, \lb{Phi}
\eqn
where $u=\pi \mathcal{M} f$ with $f=\omega/2 \pi$ being the frequency of the GWs, $\mathcal{M} = (1+z) \mathcal{M}_c$ is the measured chirp mass, and $\mathcal{M}_c \equiv (m_1 m_2)^{3/5}/(m_1+m_2)^{1/5}$  is the chirp mass of the binary system with component masses $m_1$ and $m_2$. With the above phase corrections, the waveform of the two polarizations $h_+(f)$ and $h_\times(f)$ becomes
\bqn
h_{+, \times}(f) =h_{+, \times}^{\rm GR}(f) e^{- i \delta \Psi}.  \lb{hpc}
\eqn
This expression represents the modified waveform of GWs we use to compare with the GW data.

\section {Bayesian inference and parameter estimation}
\lb{sec4}
\renewcommand{\theequation}{4.\arabic{equation}} \setcounter{equation}{0}

In this section, we describe the Bayesian inference by using observational data from LVK to constrain the coefficients describing LIV in the SME framework. Up to now, the GWTC-3 catalog contains 90 compact binary coalescence events \cite{LIGOScientific:2021djp}, including binary neutron stars (BNS) GW170817 and GW190425, neutron star–black hole binaries (NSBH), and binary black holes (BBH). Bayesian inference is an important part of modern astronomy. When we have GW data $d_i$, we compare the GW data with the predicted GW strain with LIV effects to infer the distribution of the parameters $\vec{\theta}$ which describe the waveform model. According to Bayes theorem, the posterior distribution is given by:
\bqn
 P({\vec \theta}|d,H)=\frac{P(d| \vec{\theta},H) P(\vec{\theta}| H)}{P(d|H)},
 \eqn
where $ P({\vec \theta}|d, H)$ denotes the posterior probability distributions of physical parameters ${\vec{\theta}}$ which denotes the model parameters. $H$ denotes the waveform model, $P(\vec{\theta}| H)$ denotes the prior distribution when given the model parameters $\vec{\theta}$, the denominator $P(d| \vec{\theta}, H)$ denotes the likelihood  given a specific set of model parameters and $P(d|H)$ is normalization factor called the ``evidence", 
 \bqn
 P(d|H) \equiv \int d \vec{\theta} P(d| \vec{\theta}, H) P(\vec{\theta}| H).
 \eqn

 In most cases, the GW signal is very weak and the matched filtering method can be used to extract these signals from the noises. Here, we assume that the noise is Gaussian and stationary \cite{Cutler:1994ys, Romano:2016dpx, Thrane:2018qnx}. The likelihood function of the matched filtering method can be written in the following form,
 \bqn
 P(\boldsymbol{d}|{\boldsymbol{\theta}}, H) \propto \prod_{i=1}^{n} e^{-\frac{1}{2}\langle \boldsymbol{d_i}-\boldsymbol{h({\theta})}|\boldsymbol{d_i}-\boldsymbol{h({\theta})}\rangle} ,
 \eqn
 where $\boldsymbol{h({\theta})}$ is the GW strain given by the waveform model $H$ and $i$ represents different GW detectors. The noise weighted inner product $\langle A|B \rangle$ is defined as
 \bqn
 \langle A|B \rangle = 4\; {\rm Re} \left[\int_0^\infty \frac{A(f) B(f)^*}{S(f)} df\right],
 \eqn
 where $^*$ denotes complex conjugation and $S(f)$ is the power spectral density (PSD) function of the detector. We use the PSD data encapsulated in LVK posterior sample which could lead to a more stable and reliable parameter estimation compared with obtaining the PSD from strain data by Welch averaging \cite{LIGOScientific:2018mvr, Cornish:2014kda, Littenberg:2014oda}. 

Next, we restrict our attention to the cases with LIV in the SME framework. We utilize the Python package BILBY \cite{Romero-Shaw:2020owr, Ashton:2018jfp} to perform Bayesian inference by analyzing the GW data of the 90 BBH and BNS, and NBSH merger events in the GWTC-3 catalog. We use the waveform template given in (\ref{hpc}) with (\ref{Phi}) denoting the LIV effect. We employ template \texttt{IMRPhenomXPHM} \cite{Garcia-Quiros:2020qpx, Pratten:2020ceb, Pratten:2020fqn} for the GR waveform $h^{\rm GR}_{+, \; \times}(f)$ for BBH and NSBH events, and \texttt{IMRPhenomPv2\_NRTidal} for BNS events. Since the spherical expansion coefficient formula in (\ref{zeta0}) is a general solution for different events in the same coordinate system, we can directly combine the posterior of a single event,
\bqn
P(\vec{\theta}|\{d_i\}, H) \propto \prod_{i=1}^{N} P(\vec{\theta}| d_i, H), \lb{combine}
\eqn
where $d_i$ denotes data of the $i$-th GW event and $N$ denotes selected number of the GW events.

\section {Results}
\renewcommand{\theequation}{5.\arabic{equation}} \setcounter{equation}{0}
\lb{sec5}

In this section, we present the results of the constraints on the anisotropic non-birefringent dispersion by comparing the modified waveform (\ref{hpc}) with the strain data from GW detectors. As we have mentioned, one expects the operators with the lowest mass dimension to have the dominant LIV effects on the propagation of GWs. For this reason, we are more interested in the lowest mass dimensions, for example, $d=4$ and $d=6$ for $k_{(I)jm}^{(d)}$. However, the case of $d=4$ only induces a frequency-independent effect in the modified dispersion, so they do not give any observable dephasing effects. In \cite{LIGOScientific:2017zic}, combined with the detection of the electromagnetic counterpart, the case of $d = 4$ is discussed. In this paper, we only consider the case of $d=6$.

For the case of $d=6$, the phase correction in the waveform takes the form
\bqn
\delta \Psi = A_{\bar \mu} (\pi f)^3,
\eqn
with 
\bqn
A_{\bar \mu} &=& \frac{8}{3}\zeta_{(d)}^0({\bf \hat n}) \int_{t_e}^{t_0} \frac{dt}{a^3} \nb\\
& =&  \frac{8}{3} \left(\sum_{jm}Y_{jm}({\bf \hat n}) k_{(I)jm}^{(d)}\right) \int_{t_e}^{t_0} \frac{dt}{a^3}.  \lb{Au}
\eqn
Note that $A_{\bar \mu}$ is the parameter we sampled in the Bayesian inference along with other GR parameters. We refer to the selected time interval and signal duration of PSD in \cite{Romero-Shaw:2020owr} as well as the prior selection method in Bayesian inference. And we refer to the sampling frequency and minimum frequency in Appendix E of \cite{LIGOScientific:2021djp}. 

For mass dimension $d=6$, the index $j$ can take $0, 1, 2, 3, 4$, and the index $m$ runs from $-j$ to $j$. Note that each of $k_{(I)jm}^{(6)}$ are complex functions which satisfies $k_{(I)jm}^{(6)*}=(-1)^{m}k_{(I)j-m}^{(6)}$. Thus the number of independent components for coefficients $k_{(I)jm}^{(6)}$ are $(d-1)^2=25$. The number of independent components can refer to Ref. \cite{Kostelecky:2016kfm}. These components are entirely tangled together. It is prohibitively that one can break the degeneracy of the coefficient $k_{(I)jm}^{(6)}$ by sufficient GW events since each event has different source locations. Here we adopt another approach by using the ``maximum-reach" method, with which one can constrain each of these components separately \cite{Shao:2020shv, Wang:2021ctl, Niu:2022yhr}. This implies that when one considers one of these components, the others are set to zero. It is worth mentioning here that in \cite{Shao:2020shv, Wang:2021ctl}, an attempt to place global constraints is proposed, which can place limits on all components simultaneously. 
However, since there are some non-Gaussian features in our posteriors, it is not appropriate to directly follow that method where the possible time delays are assumed to be Gaussian and the multi-Gaussian likelihood as a function of all coefficients can be constructed. This issue currently requires further research, we leave it to future works. This study only reports the results of the ``maximum-reach" method rather than global constraints. 

Data samples of $A_{\mu}$, right ascension (ra), and declination (dec) were obtained by Bayesian reference, so each component of $k_{(I)jm}^{(6)}$ characterizing LIV effects can be calculated from the posterior samples of $A_\mu$, ra, and dec via Eqs. (\ref{Au}) and (\ref{zeta}). With a fixed coordinate system, the expansion coefficients are supposed to be the same for all events. We combined the individual posterior samples of $k_{(I)jm}^{(6)}$ through (\ref{combine}), and the results were shown in Fig. \ref{violin_pdf}. Table \ref{jm_const} summarized the $90\%$ confidence interval of each LIV coefficents $k_{(I)jm}^{(6)}$. Fig. \ref{violin_pdf} shows that most of the expansion coefficients $k_{(I)jm}^{(6)}$ are roughly on the same order of magnitude, although the magnitude of $k^{(6)}_{({I})33}$ and $k^{(6)}_{({I})44}$ coefficients is a little bit larger. From both the Fig.~\ref{violin_pdf} and Table.~\ref{jm_const}, it is obvious that the posterior samples and the $90\%$ confidence interval of each coefficient $k_{(I)jm}^{(6)}$ are all consistent with zero, which indicates there are no any signatures of the LIV arising in the linearized gravity of SME has been found in the GW signals.

\section{conclusion}
\renewcommand{\theequation}{6.\arabic{equation}} \setcounter{equation}{0}
\lb{sec6}

Since the detection of GW signals by LIGO/Virgo Collaboration, the tests of gravity in the strong field regime with GWs have become possible. With the increase in detector number and sensitivity, LVK catalog GWTC-3 now contains 90 GW events. In this paper, we consider the gauge-invariant linearized gravity sector of SME to investigate the Lorentz-violating effects in the propagation of GWs and constrain them with the detected GW data. The Lagrangian in the SME framework contains all possible gauge-invariant quadratic terms of the metric perturbation $h_{\mu \nu}$, which represent the LIV modification. According to a similar approach developed in the discussion of Lorentz symmetry in the photon sector of SME \cite{Kostelecky:2009zp}, a modified dispersion relation of GWs can be obtained from the Lagrangian, in which the LIV effects without birefringence is completely characterized by the coefficient $k_{(I)jm}^{(d)}$. When $j \neq 0$, $k_{(I)jm}^{(d)}$ is direction-dependent, which can lead to anisotropic effects in the propagation of GWs. We derive the modified waveform of GWs with the LIV effects and performed Bayesian inference on the GWs data to constrain these effects.  

By comparing the modified waveform (\ref{hpc}) with the strain data from the GWTC-3 catalog, we derive the constraints on the effects of the anisotropic non-birefringent dispersions of GWs due to the LIV in the gauge invariant linearized gravity sector of the SME. As we mentioned, we expect the operators with the lowest mass dimension in the SME to have a major LIV effect on the propagation of GWs. Therefore, we are more interested in the lowest dimensions, for example, for $k_{(I)jm}^{(d)}$, one has $d = 4$ and $d = 6$. However, when $d = 4$, only frequency-independent effects are produced in the modified dispersion relation, so it cannot give any observable out-of-phase effects. But in \cite{LIGOScientific:2017zic}, combined with the detection of the electromagnetic counterpart, the case of $d = 4$ is discussed. In this paper, we only consider the case of $d=6$. Here we use the ``maximum-reach" method to constrain each component of $k_{(I)jm}^{(d)}$ separately. 

\begin{figure*}
    \centering
    \includegraphics[width=16cm]{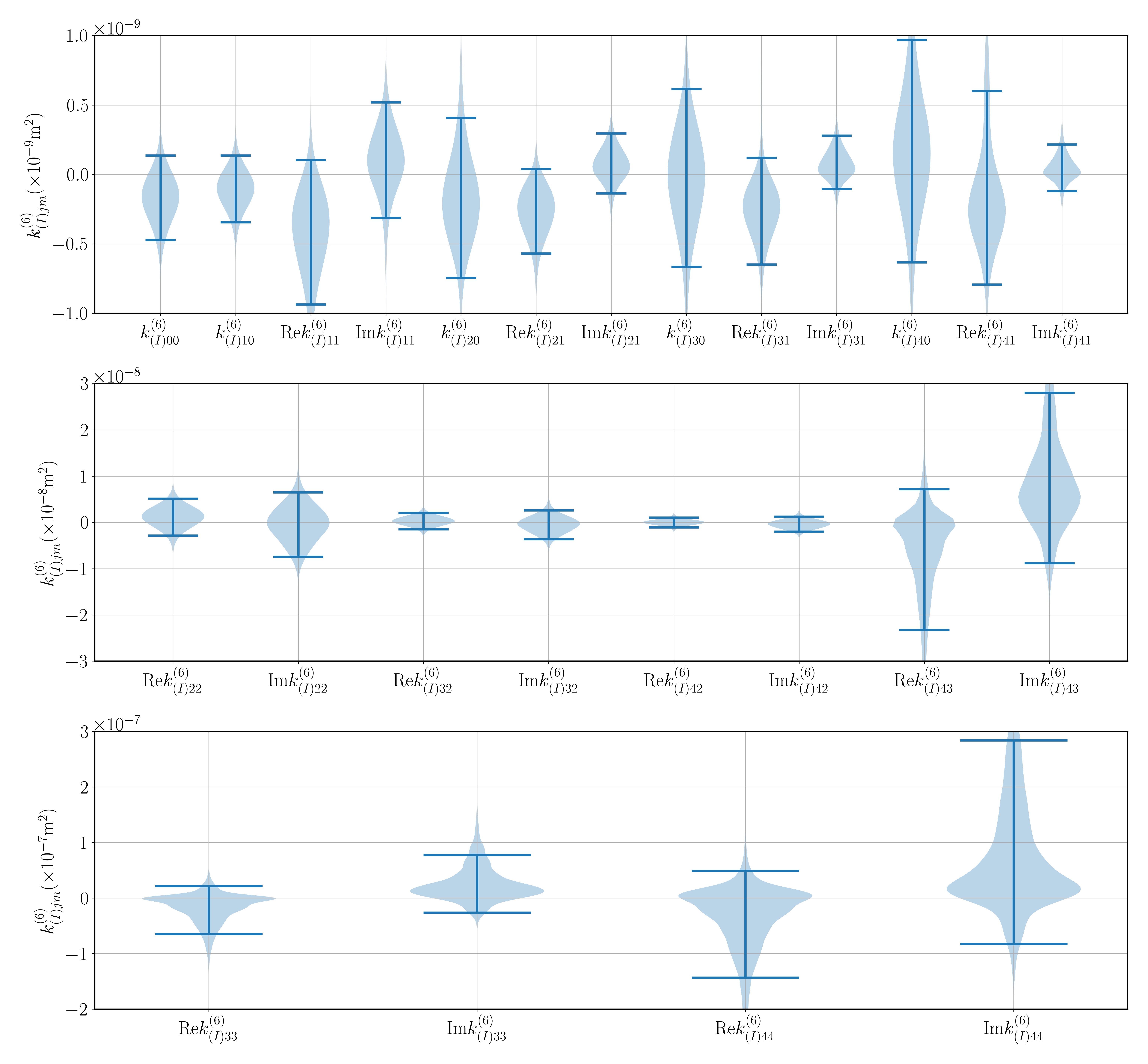}
    \caption{Combined probability distributions of each component of $k_{(I)jm}^{(6)}$. We have drawn 25 violin plots of $k_{(I)jm}^{(6)}$ components and the error bars denote the $90\%$ confidence intervals, whose central value is basically around zero, consistent with GR. 
   }
    \label{violin_pdf}
\end{figure*}

Results represented in Fig. \ref{violin_pdf} show that there is no evidence of any violation of Lorentz symmetry. Therefore, we give the constraints of the coefficients $k_{(I)jm}^{(6)}$ describing anisotropic non-birefringent effects. Most of the $k_{(I)jm}^{(6)}$ components have a $90\%$ confidence interval of roughly $10^{-10}$, with some $k_{(I)jm}^{(6)}$ components having a $90\%$ confidence interval of between $10^{-8}$ and $10^{-7}$, but the medians for all of these components is around zero. Constrains on each component of $k_{(I)jm}^{(6)}$ are summarized in table \ref{jm_const}. Since the next generation of GW detectors can detect lighter and more distant BBH and BNS events, it is expected such systems can lead to tighter constraints on non-birefringent dispersions in the future.

\begin{table}
\caption{$90\%$ confidence interval of each component of the LIV coefficients $k_{(I)jm}^{(6)}$ from 90 GW events in the GWTC-3 catalog.}
\label{jm_const}
\centering  
\begin{ruledtabular}
    \begin{tabular}{cc|cc}
    $j$ & $m$ & Coefficient  &  Constraint ($10^{-9} \; {\rm m}^2$)\\
    \hline
    $0$ & $0$ & $k^{(6)}_{(IV)00}$ & $(-0.5,0.1)$\\
    $1$ & $0$ & $k^{(6)}_{(IV)10}$ & $(-0.3,0.1)$\\
        & $1$ & Re   $k^{(6)}_{(IV)11}$ & $(-0.9,0.1)$\\
        &     & Im   $k^{(6)}_{(IV)11}$ & $(-0.3,0.5)$\\
    $2$ & $0$ & $k^{(6)}_{(IV)20}$ & $(-0.7,0.4)$\\
        & $1$ & Re   $k^{(6)}_{(IV)21}$ & $(-0.6,0.0)$\\
        &     & Im   $k^{(6)}_{(IV)21}$ & $(-0.1,0.3)$\\ 
        & $2$ & Re   $k^{(6)}_{(IV)22}$ & $(2.9,5.1)$\\
        &     & Im   $k^{(6)}_{(IV)22}$ & $(-7.4,6.5)$\\
    $3$ & $0$ & $k^{(6)}_{(IV)30}$ & $(-0.7,0.6)$\\
        & $1$ & Re   $k^{(6)}_{(IV)31}$ & $(-0.6,0.1)$\\
        &     & Im   $k^{(6)}_{(IV)31}$ & $(-0.1,0.3)$\\
        & $2$ & Re   $k^{(6)}_{(IV)32}$ & $(-1.5,2.1)$\\
        &     & Im   $k^{(6)}_{(IV)32}$ & $(-3.6,2.6)$\\
        & $3$ & Re   $k^{(6)}_{(IV)33}$ & $(-64.9,21.6)$\\
        &     & Im   $k^{(6)}_{(IV)33}$ & $(-26.4,77.7)$\\     
    $4$ & $0$ & $k^{(6)}_{(IV)40}$ & $(-0.6,1.0)$\\
        & $1$ & Re   $k^{(6)}_{(IV)41}$ & $(-0.8,0.6)$\\
        &     & Im   $k^{(6)}_{(IV)41}$ & $(-0.1,0.2)$\\
        & $2$ & Re   $k^{(6)}_{(IV)42}$ & $(-1.0,1.0)$\\
        &     & Im   $k^{(6)}_{(IV)42}$ & $(-2.0,1.2)$\\
        & $3$ & Re   $k^{(6)}_{(IV)43}$ & $(-23.2,7.2)$\\
        &     & Im   $k^{(6)}_{(IV)43}$ & $(-8.8,28.0)$\\
        & $4$ & Re   $k^{(6)}_{(IV)44}$ & $(-143.3,48.9)$\\
        &     & Im   $k^{(6)}_{(IV)44}$ & $(-82.5,284.3)$\\
    \end{tabular}
    \end{ruledtabular}
\end{table}

\section*{Acknowledgements}
T.Z., Q.W., and A.W. are supported in part by the National Key Research and Development Program of China under Grant No. 2020YFC2201503, the Zhejiang Provincial Natural Science Foundation of China under Grants No. LR21A050001 and No. LY20A050002, the National Natural Science Foundation of China under Grants No. 12275238, No. 11975203, No. 11675143, and the Fundamental Research Funds for the Provincial Universities of Zhejiang in China under Grant No. RF-A2019015.
W.Z. is supported by the National Key Research and Development Program of China (Grant No. 2022YFC2204602 and 2021YFC2203102) and the National Natural Science Foundation of China (Grants No. 12273035), the Fundamental Research Funds for the Central Universities. 
X.Z. is supported by the National Natural Science Foundation of China (Grants No. 11975072 and No. 11835009) and the National Key Research and Development Program of China (Grants No. 2022SKA0110200 and No. 2022SKA0110203). We are grateful that this research has made use of data or software obtained from the Gravitational Wave Open Science Center (gw-openscience.org), a service of LIGO Laboratory, the LIGO Scientific Collaboration, the Virgo Collaboration, and KAGRA. 

\appendix

\end{document}